\documentclass{llncs}

\usepackage{microtype}%if unwanted, comment out or use option "draft"
\usepackage{float}
\usepackage{latexsym}
\usepackage{enumitem}
\usepackage[final]{pdfpages}        % Include pages from another PDF document
\usepackage{xcolor}
\usepackage{listings}
\usepackage{placeins}

\usepackage{amsmath}
\usepackage{amssymb}
\usepackage{listings}  
\usepackage{stmaryrd}
\usepackage{graphicx}
\usepackage{bbm,relsize}
\usepackage{mathbbol}
\usepackage{mathabx}
\usepackage{calrsfs}
\usepackage{url}
\usepackage{euscript}

\definecolor{pblue}{rgb}{0.13,0.13,1}
\definecolor{pgreen}{rgb}{0,0.5,0}
\definecolor{pred}{rgb}{0.9,0,0}
\definecolor{pgrey}{rgb}{0.46,0.45,0.48}

\renewcommand{\&}{\binampersand}
\newcommand{\zero}{{\sf 0}}

\renewcommand{\root}[1]{{\it root}(#1)}
\newcommand{\typeof}[2]{{\it typeof}(#1,#2)}

\newcommand{\nulam}[1]{(\nu \, #1)}
\newcommand{\subst}[2]{\{\raisebox{.5ex}{\footnotesize$#1$}  /
                       \raisebox{-.5ex}{\footnotesize$#2$} \}}

\newcommand{\name}[1]{{\it names}(#1)}

\newcommand{\lock}[1]{\textit{lock}_{#1}}

\newcommand{\tset}{T}
\newcommand{\rset}{R}

\newcommand{\set}[1]{#1}

\newcommand{\recty}{\rho}

\newcommand{\destr}[1]{{\it env}(#1)}

\newcommand{\constr}[1]{{\it mk\_tree}(#1)}

\newcommand{\LAM}{\mathcal{L}}

\newcommand{\bigfract}[2]{\frac{^{\textstyle #1}}{_{\textstyle #2}}}

\newcommand{\void}{{\it void}}

\def \mathrule #1#2#3{ {\footnotesize 
                       \bigfract{#2}{#3}
                       }
                      }

\newcommand{\Java}{{\tt Java}}
\newcommand{\JVM}{{\tt JVM}}
\newcommand{\JVMLd}{\mbox{\tt{JVML}$_d$}}
\newcommand{\JVML}{\mbox{{\tt JVML}}}
\newcommand{\JaDA}{\texttt{JaDA}}

\newcommand{\C}{{\tt{C}}}
\newcommand{\T}{{\tt{T}}}
\newcommand{\D}{{\tt{D}}}

\newcommand{\E}{{\tt{E}}}
\newcommand{\FD}{\mbox{\textit{FD}}}
\newcommand{\MD}{\mbox{\textit{MD}}}

\newcommand{\CF}{\mbox{\textit{CF}}}
\newcommand{\Int}{\mbox{\texttt{int}}}
\newcommand{\ty}{{\tt{T}}}
\newcommand{\m}{{\tt{m}}}
\newcommand{\f}{{\tt f}}
\newcommand{\g}{{\tt g}}
\newcommand{\vect}[1]{\overline{#1}}

\newcommand{\tpair}[3]{(#2,#3)_{#1}}
\newcommand{\aand}{\, \&\, }

\newcommand{\ztop}[1]{\lceil #1 \rceil}

\newcommand{\field}[3]{#1.#2:#3}

\newcommand{\att}[1]{#1_{\mbox{{\scriptsize {\tt \$}}}}}

\newcommand{\invk}[3]{#1.#2(#3)}

\newcommand{\new}{\mbox{{\tt new}}}
\newcommand{\putf}{\mbox{{\tt putfield}}}
\newcommand{\getf}{\mbox{{\tt getfield}}}
\newcommand{\monitorenter}{\mbox{{\tt monitorenter}}}
\newcommand{\monitorexit}{\mbox{{\tt monitorexit}}}
\newcommand{\return}{\mbox{{\tt return}}}

\newcommand{\start}{\mbox{{\tt start}}}
\newcommand{\ifL}[1]{\mbox{{\tt if}}~#1}
\newcommand{\invokevirtual}[1]{\mbox{{\tt invokevirtual}}~#1}

\newcommand{\inc}{\mbox{\texttt{inc}}}
\newcommand{\pop}{\mbox{\texttt{pop}}}
\newcommand{\load}{\mbox{\texttt{load}}}
\newcommand{\store}{\mbox{\texttt{store}}}
\newcommand{\goto}{\mbox{\texttt{goto}}}

\newcommand{\push}{\mbox{\texttt{push}}}
\newcommand{\Addr}{\mbox{\textsc{Addr}}}

 \newcommand{\lam}{\ell}
 \newcommand{\cct}{\mbox{\textsc{bct}}}

 \newcommand{\of}[2]{{#1_{#2}}}

\newcommand{\dom}[1]{{\it dom}(#1)}

\title{Deadlock detection of Java Bytecode}
\author{Abel Garcia and Cosimo Laneve}
\institute{Dept.~of Computer Science and Engineering, University of Bologna -- INRIA Focus}

\begin{document} 

\maketitle
\begin{abstract}
This paper presents a  technique for deadlock detection of {\Java} programs.
The technique uses typing rules for extracting infinite-state abstract models of the 
dependencies among the components of the {\Java} intermediate
language -- the {\Java} bytecode.
Models are subsequently analysed by means of an extension of a 
solver
that we have defined for detecting deadlocks in process calculi.
Our technique is complemented by a prototype verifier that also 
covers most of the {\Java} features.
 \end{abstract}
%
%\category{F.3.1}{Logics and meanings of programs}{Specifying and Verifying and Reasoning about Programs}[Mechanical verification]
%\category{F.3.2}{Logics and meanings of programs}{Semantics of Programming Languages}[Operational semantics,Program analysis ]
%\category{F.1.1}{Computation by abstract devices}{Models of Computation}[Relations between models]
%
%
%\terms
%Verification of program properties, deadlock analysis, {\Java} bytecode.
%
%
%\keywords
%....
%%concurrent object-oriented programming.

%%%%%%%%%%%%
\section{Introduction}
\label{sec:intro}
%!TEX root = LOPSTR2017.tex

Deadlocks are common flaws of concurrent programs that
occur when a set of threads are blocked because each one is attempting to 
acquire a lock held by another one. Such 
errors are difficult to detect or anticipate, since they may not happen 
during every execution, and may have catastrophic effects for the overall 
functionality of the software system. 
At the time of writing this paper,  the 
\emph{Oracle Bug Database}\footnote{http://bugs.java.com/} reports more 
than 40 unresolved bugs due to deadlocks, while the 
\emph{Apache Issue Tracker}\footnote{https://issues.apache.org/jira} 
reports around 400 unresolved deadlock bugs. These two databases refer to 
programs 
written in {\Java}, a mainstream programming language in a lot of domains, 
such as web and cloud applications, user applications and mobile applications. 

The objective of our research is to design and implement a technique capable of detecting potential deadlock bugs of {\Java} programs \emph{at static time}. %
This objective is difficult because  {\Java} has a complex concurrency model:
it uses threads that may perform 
read/write operations over shared variables and whose execution 
depends on the 
scheduling strategy implemented in the Java Virtual Machine ({\JVM}). 
In addition, {\Java}, 
being a full-fledged programming language, includes an extensive standard library with lots of features implemented in native language.

To reduce the complexity of our work, we decided to address the  {\Java} 
bytecode, namely 198 instructions that are the compilation target of 
every {\Java} application and have a reference semantics that 
is defined by the {\JVM} behaviour. Therefore, it is possible to deliver
correctness results without narrowing/oversimplifying  our original goal. 
In this paper, we present our technique on a
subset of {\Java} bytecode, called {\JVMLd}, which
includes basic instructions for concurrency, such as 
thread creations, synchronizations, and creations of
new objects. The language is defined in Section~\ref{sec:lang}.

The technique consists of two stages. 
The first stage defines
a type system that reconstructs the concurrent behaviour of methods.
The key principles are the following ones.
Each method has an associated type that depends on the type of the 
arguments (the object ``{\tt this}'' is one argument) and that 
expresses the 
\emph{concurrent behaviour}.  
This ``concurrent behaviour'' 
reports (\emph{i}) the 
 \emph{sequence of locks} that has been acquired/released by the method,
(\emph{ii})
the \emph{threads created}, and (\emph{iii})
the \emph{methods that have been invoked}. It includes the analysis of aliases
that traces the creation of new objects and their copies 
(because {\JVMLd} instructions may create and copy objects).
The alias analysis is performed in a \emph{symbolic way} by using a 
\emph{finite set of names}: this is a critical part of our technique 
because methods may create threads and,
when methods are either recursive or iterative, the set of created threads  may be 
infinite. In particular, we had to devise finite representatives of
(infinite sets of) thread names that are sound with respect to the 
(deadlock) analysis.
Section~\ref{sec:examples} reports a code that can be 
written in (a simple extension of) {\JVMLd} and that is problematic as
regards  deadlock detection. 
Section~\ref{sec:analysis} describes the type system and Section~\ref{sec.moreandJaDA} overviews the typing of complex features of {\JVML}.

The second stage of our technique defines the analysis 
of the behavioural model.
In fact, the three reports above --  (\emph{i}), (\emph{ii}), and 
(\emph{iii}) -- are terms in a modelling language that extend so-called
\emph{lams}~\cite{GiachinoL14,GKL2014,KobayashiLaneve}. Lams are 
conjunctions and disjunctions of object dependencies and method invocations and
the extension has been necessary for modelling {\Java} \emph{reentrant
locks}. In particular, our dependencies also carry thread names -- 
$\tpair{t}{a}{b}$ means that the thread $t$, which owns the lock
of $a$, is going to lock $b$. In {\Java}, the lam $\tpair{t}{a}{a}$ is not a 
circular dependency because it means that 
$t$ is acquiring  the same lock \emph{twice}. Because of this extension,
the algorithm for detecting circularities in 
lams is different than the one in~\cite{GKL2014,KobayashiLaneve}.
We address this issue in  
Section~\ref{sec:algorithm}.

Our deadlock detection  technique has been prototyped and the verifier is called
{\JaDA}. 
While the type system in 
this paper simply checks static information,  
{\JaDA} infers
the behavioural types from the
bytecode.
Inference is important in 
practice because it lightens the analysis but checking is crucial 
for type safety~\footnote{The technical details of type safety appear in the full paper, where we also overview the inference system of {\JaDA}.}.
{\JaDA} includes several features of {\JVML}; 
this has made possible to
deliver initial assessments of the tool, which are discussed in 
Section~\ref{sec.JaDA}. Section~\ref{sec:conclusions}
discusses
related work and reports our concluding remarks.

%

%%%%%%%%%%%%
\section{Overview of {\JVML} and of our technique}
\label{sec:examples}
%!TEX root = LOPSTR2017.tex

Figure~\ref{fig:buildNetwork} reports a {\Java} class
called {\tt Network} and some of its {\JVMLd} representation.
The corresponding {\tt main} method creates a network of {\tt n} threads -- the 
philosophers --
by invoking  {\tt buildNetwork} -- say $t_1, \cdots , t_n$ -- 
that are all potentially running in parallel with the caller -- say $t_0$.
Every two adjacent philosophers share an object -- the fork --, which is also created 
by {\tt buildNetwork}.
\noindent\begin{figure}[t]
{\scriptsize
\begin{minipage}[t]{0.40\linewidth}
\begin{verbatim}
class Network{

 public void main(int n){
   Object x = new Object();
   Object y = new Object();
   buildNetwork(n, x, y); //no deadlock
   buildNetwork(n, x, x); // deadlock
 }
 
 public void buildNetwork(int n, 
                Object x, Object y){
   if (n==0) { 
     takeForks(x,y) ; 
   } else { 
     final Object z = new Object() ;
     Thread t = new Thread(){
       public void run(){ 
         takeForks(x,z) ; 
     }} ;		
     t.start(); 
     this.buildNetwork(n-1,z,y) ;
   }
 }

 public void takeForks(Object x, 
                       Object y){
   synchronized(x){ synchronized(y){ } }
 } 
}
\end{verbatim}
\end{minipage}
\begin{minipage}[t]{0.6\linewidth}
\begin{verbatim}
  public void buildNetwork(int n, Object x, Object y)
     0  iload_1          //n
     1  ifne 13
     4  aload_0          //this
     5  aload_2          //x
     6  aload_3          //y
     7  invokevirtual 24 //takeForks(x, y):void
    10  goto 50
    13  new 3 
    16  dup
    17  invokespecial 8  //Object() 
    20  astore 4         //z
    22  new 26 
    25  dup
    26  aload_0          //this
    27  aload_2          //x
    28  aload 4          //z
    30  invokespecial 28 //Network$1(this, x, z)
    33  astore 5         //thr
    35  aload 5          //thr
    37  invokevirtual 31 //start():void 
    40  aload_0          //this
    41  iload_1          //n
    42  iconst_1
    43  isub
    44  aload 4          //z
    46  aload_3          //y
    47  invokevirtual 36 //buildNetwork(n-1, z, y):void 
    50  return
\end{verbatim}
\end{minipage}
}
\vspace*{-.2cm}
\caption{Java Network program and corresponding bytecode (only the {\tt buildNetwork} method). 
}
\label{fig:buildNetwork}
\vspace*{-0.5cm}
\end{figure}
Every thread $t_i$ locks the two adjacent forks, that 
are passed as (implicit) arguments of the thread, and terminates -- this is 
performed by the method {\tt takeForks}. 
It is well-known that when the network is a table (it is circular -- the thread $t_n$ is sharing 
one of its forks with $t_0$) and  all the threads have a symmetric strategy of locking 
objects then a deadlock may occur. On the contrary, when either the network is
not circular or one thread has 
an anti-symmetric strategy, no deadlock will ever occur. 
%%%
Therefore {\tt buildNetwork(n,x,y)} is deadlock free, while 
{\tt buildNetwork(n,x,x)} is deadlocked (when ${\tt n}>0$).

The problematic issue of {\tt Network} is that the number of threads
 is not known statically because {\tt n} is an argument of 
{\tt main}. This is displayed in the bytecode of {\tt buildNetwork} in
Figure~\ref{fig:buildNetwork} by the instruction at address 30 where a 
new thread is created and by the instruction at address 37 where the thread is started.
The recursive invocation that causes the (static) unboundedness is found at 
 instruction  47.
Our technique is powerful enough to cope with such 
problems and to predict the correct behaviour of the code of Figure~\ref{fig:buildNetwork}
and the faulty one if we comment {\tt buildNetwork(n,x,y)} and de-comment 
{\tt buildNetwork(n,x,x)}.
The technique works as follows. It infers abstract methods' behaviors 
by computing types, called \emph{lams}, of their  bytecode bodies. 
These lams abstract each bytecode instruction by dropping the \emph{non-relevant} 
information for the deadlock analysis (e.g.~operations on integer variables). 
In practice, the relevant operations for deadlock analysis are: locking operations ({\tt monitorenter} and {\tt monitorexit} instructions), thread spawning operations, function invocations and objects' structures. Thereafter
the abstract model is analysed by a solver.

% 

%%%%%%%%%%%% 
\section{The language {\JVMLd}}
\label{sec:lang}
%!TEX root = LOPSTR2017.tex

{\JVMLd} is a restriction of {\JVML} that includes basic constructs and instructions for concurrency~\footnote{Actually,
{\JVMLd} has a minor difference with respect to {\JVML}:
in {\JVML}, local variables are addressed by non-negative integers instead
of names. 
}. In {\JVMLd}, a program is a collection of 
\emph{class files} whose methods have bodies written in {\JVMLd} bytecode.
This bytecode is a partial map from \emph{addresses} {\Addr} to instructions.
Addresses, ranged over by $L$, $L'$, $\cdots$, are 
intended to be natural numbers and we  use the 
function $L+1$ that returns the least address that is strictly greater than $L$.
When $P$ is a program, we write $\dom{P}$ to refer to its domain (the set
of addresses) and we assume that $0 \in \dom{P}$ for every bytecode $P$.

We  use a number of \emph{names}: for classes, ranged over by $\C$, 
$\D$, $\cdots$, for fields, ranged over by $\f$,
$\f'$, $\cdots$, for methods, ranged over by $\m$, $\m'$, $\cdots$, and for 
local variables, ranged over by $x$, $y$, $\cdots$. 
A possible empty sequence of names or syntactic categories of the following 
grammar is written by 
over-lining the name or the syntactic category, respectively. 
For instance 
a sequence of local variables is written 
$\vect{x}$. However, when we need to access to the elements of a sequence, 
we use the 
notation $x_1, \cdots, x_n$.
Class files {\CF} are defined by the grammar:
\[
\begin{array}{lcl}
  \CF &::=& class~\C~\{  
  fields: \overline{\FD} ~~ methods: \overline{\MD} \} \\
  \FD &::=& \field{\C}{\f}{\ty} 
  \end{array}
  \qquad 
\begin{array}{lcl} 
  \MD &::=& \ty ~\m~(\C,\overline{\ty})~P 
  \\
  \ty&::=& \top ~|~ \Int ~|~ \C 
\end{array}
\]
where ``${\it fields}:$'' and ``${\it methods}:$'' are keywords
and $\top$ is a special type that include all the other types 
(any value of any type has 
also type $\top$). This type will represent values that are unusable in our static
semantics. The type name $\C$ represents a class type, \emph{which is
never recursive} in {\JVMLd}.

Instructions $\mathit{Instr}$ of {\JVMLd} bytecode are
of the following form:
\[
\begin{array}{rl}
\! \! \mathit{Instr} ::=& \inc ~|~ \pop ~|~ \push ~|~ \load~x ~|~ \store~x ~|~ 
                     \ifL{L} ~|~ \goto\,L  
 \\ | & \new\,\C ~|~  \putf\,\field{\C}{\f}{\ty} ~|~ \getf\,\field{\C}{\f}{\ty} 
 ~|~ \monitorenter ~|~\monitorexit
 \\
   | & \invokevirtual{\invk{\C}{\m}{\overline{\ty}}}
    ~|~ \start~\C  %~|~ \join~\C
  ~|~  \return
\end{array}
\]
The informal meaning of these instructions is as follows:
\begin{itemize}
\item  $\inc$ increments the content of the stack; $\pop$ and $\push$,
  respectively, pops an element from the stack and pushes the integer 0 on the stack; 
  $\load~x$ and $\store~x$ respectively loads the value of $x$ on the stack and
  pops the  top value of the stack by storing it in $x$; 
  $\ifL{L}$ pops
  the top value of the stack and either jumps to the instruction
  at address $L$, if it is nonzero, or goes to the next instruction;
  $\goto\,L$ is the 
  unconditional jump;
  \item 
  $\new\,{\C}$
  allocates a new object of type $\C$, initializes it and pushes it on top
  of the stack; $\putf\,\field{~\C}{\f}{\ty}$ pops the value on the stack and the
  underlying object value, and assigns the former to the field $\f$ of the
  latter; $\getf\,\field{~\C}{\f}{\ty} $ pops the object on the stack and pushes the
  value in the field $\f$ of that object;

   \item $\monitorenter$, $\monitorexit$ are the synchronization primitives that 	pop the object on the stack and lock and unlock it, respectively;

  \item $\invokevirtual{\invk{\C}{\m}{\ty_1, \cdots , \ty_n}}$ pops $n$ values from
  the stack (the arguments of the invocation) and dispatches the method $\m$ on the
  object on top of the stack; when the method terminates, the returned value is pushed 
  on the stack;
  
   \item $\start~\C$ creates and starts a new
  thread for the object on top of the stack.
  This operation corresponds to {\tt invokevirtual java/lang/Thread/start()} 
   on a thread of class $\C$
  in {\JVML}. We separate it from {\tt invokevirtual} in order to
  provide more structure to our semantics
 (because it has an effect on the set of threads -- see the operational
  semantics in the Appendix, where we also consider the instruction {\tt join});
 
\item $\return$ terminates program execution.  

\end{itemize}
\noindent
The bytecode in Figure~\ref{fig:buildNetwork} is written in 
a sugared extension of {\JVMLd}. In particular, {\tt aload} and {\tt iload}
correspond to our {\tt load} instruction (when the argument is an object or
an integer, respectively), {\tt ifne} corresponds to {\tt if}, 
{\tt dup} duplicates the top of the stack, {\tt sub} subtracts the element on top 
of the stack from the last-by-one,
{\tt invokespecial} is the method invocation of the constructor of the class.

In order to simplify the presentation, in this paper we assume that fields are read-only as they cannot be modified 
after the initialisation (which is done by constructors that, in turn, are
sequential)~\footnote{The full paper 
reports the complete analysis that also addresses race conditions.}.

%

%%%%%%%%%%%%
\section{The type system}
\label{sec:analysis}
%!TEX root = LOPSTR2017.tex

The purpose of the type system is to associate lams to {\JVMLd} bytecodes.
Since {\JVMLd} is the target (of large part) of {\Java}, the association is complex
because we must deal with objects and aliasing, object creation and updates performed
by constructors, and the concurrent operations
-- creation of new threads, lock and unlock operations. 
Therefore the details are pretty technical. In this paper we
overview the type system by discussing features that are increasingly difficult.
In particular, we will discuss one typing rule -- 
that of {\tt invokevirtual} -- and we will study the basic, sequential case, the
case of invocation of constructors, and the case of invocation of a concurrent
thread. The complete set of rules appears in the full paper.

Typing rules associate types, which are \emph{lams}, 
to {\JVMLd} instructions 
by means of \emph{judgments}. 
Typically, these judgments are abstractions of the machine states. In 
case of {\JVM}, the state is
a memory, called \emph{heap} and the set of running \emph{threads}. 
In turn, every thread is a stack of activation records -- each one containing
the address of the instruction to be performed, 
a stack, and a local memory -- plus the sequence of locks owned. 
For example, in the case of {\tt invokevirtual} (without arguments), the 
the element on the stack is the called object (which is used by the 
{\JVM} to locate the right method body). 

A possible judgment for the instruction at address $i$ of the {\JVMLd} program
$P$ is
\[
\Gamma, F,S,Z,i \vdash_t P: \lam
\]
where $\Gamma$, called \emph{environment},  is the abstraction of the heap,
$F$ and $S$ are the abstraction of the local memory and the stack, 
respectively, and
$Z$ is the sequence of locks acquired by the thread. The term $\ell$ 
is the \emph{lam} of the instruction $i$ and
$t$ is a symbolic name identifying the thread that is executing the
instruction. (At static time it is not possible to model the stack of 
activation records).

Environments $\Gamma$, memories $F$, stacks $S$, and sequences of locks $Z$
are defined by means of \emph{types}, which are not lams. Types in judgments are more descriptive than those in {\JVMLd} syntax; in particular
object types are not just classes $\C$, that is 
records $[\f_1 : \T_1, \cdots, \f_n:\T_n]$, where $\f_i$ are the fields 
of the class. In fact, this notation is not adequate for dealing with aliasing.
For example, let $\C$ be a class with two fields $\f_1$ and $\f_2$ that store
objects of class $\D$. If $\C$-objects are represented by
 $[\f_1:\D, \f_2:\D]$ then it is not possible to recover the identities of
the values of $\f_1$ and $\f_2$; therefore we cannot 
distinguish the cases when $\f_1$ and $\f_2$ store the same 
object or two different objects, which is 
sensible when we compute object dependencies. 

Therefore we decided to use \emph{symbolic names}, ranged over by
$a$, $b$, $\cdots$, which also include $\void$ and \emph{thread names} 
(threads are objects in {\Java}; we use $t$, $t'$, 
$\cdots$ when a name addresses a thread). Symbolic names allow us to 
define \emph{flattened types} such as 
$[\f_1:b, \f_2:b]$ and $[\f_1:b, \f_2:c]$, thus separating the two
foregoing cases. Actually, in
order to avoid ambiguities with different classes having same field names,
the flattened types also carry the class name, 
e.g.~$([\f_1:b, \f_2:b], \C)$.

The binding of symbolic names and flattened types is defined by the
\emph{environments}, ranged over by $\Gamma, \Gamma_i, \cdots$. For example
$[a \mapsto ([\f:b], \set{\C}), b \mapsto ([\g:\Int], \D)]$ is an environment
that defines the names $a$ and $b$. 
The function $\typeof{\Gamma}{a}$ returns the type of $a$ 
in $\Gamma$. 

Finally, our type system uses \emph{vectors} $\Gamma, F,S,Z$ that 
are  indexed by the addresses 
in $\dom{P}$. The elements of these vectors are  

\begin{itemize}
\item  
$\Gamma_i$ is the environment at address $i$;
\item the map 
$\of{F}{i}$ maps local variables to type values;

\item $\of{S}{i}$ is a sequence of type values;
\item $\of{Z}{i}$ is the \emph{sequence of symbolic names} locked at
address $i$.
\end{itemize}

\paragraph{Simple methods.}
We begin with the rule for {\tt invokevirtual} of 
a method that has no argument,
does not modify the carrier and returns $\void$:
\[
\mathrule{t-invk}{
	\begin{array}{c}  
    P[i]= \invokevirtual~{\C.\m}~( ~ )\qquad
	i+1 \in \dom{P} \\
    \of{S}{i} = a  \cdot S'
	\quad
	\typeof{\of{\Gamma}{i}}{a} = \C
	\\
	\of{\Gamma}{i+1} = \of{\Gamma}{i} 	\quad		
	\of{S}{i+1} =  \void \cdot S'
	\quad
    \of{F}{i+1}= \of{F}{i}
    \quad     \of{Z}{i+1} = \of{Z}{i}
	\end{array}
	}{ 
	   \Gamma, F,S,Z,i \vdash_t P: 
       \C.\m(a,t,\ztop{\of{Z}{i}}) 
       }
\]
The rule verifies that the top element of the stack is of type $\C$ and
constraints the stack $\of{S}{i+1}$ to be the same as $\of{S}{i}$, 
except for the top element, which is replaced by ${\it void}$. The
lam of the instruction $i$ indicates that the instruction is a method
invocation: we will discuss this term later; we just notice that 
$\ztop{\of{Z}{i}}$ is the first name in the sequence $\of{Z}{i}$
(that represents the last object locked by $t$).

\paragraph{Constructors.}
We continue by analysing methods that update the carrier, such as 
constructors, and return an object. It is usual in {\Java} that the
returned object is new (in the {\JVM} it is a fresh run time name). 
In the type system we must be careful with such names.
Overall the set of symbolic names used by the type system must be finite,
which is an issue when a new object is created inside an iteration or 
a recursion. To overcome this issue, we let symbolic name represent
infinitely many instances of names in these cases. Technically, we use a 
a function $\name{ i }$ that
takes an address $i$ and returns a 
tuple of names whose length is finite and depends on the address. 
This function returns a name that may occur already in the judgment
 when the instruction is not 
executed  once.

The above rule has no element specifying (\emph{i}) the type of the returned 
object and (\emph{ii}) the number of fresh names created by the invocation.
As regards (\emph{i}), one could think to extend $\Gamma$ with the 
typing of methods. Again, this is not adequate because of our need to trace
the identity of objects. For example, let $\C.\m$ be a method that returns 
an object of class $\C$; we must distinguish
the cases when $\C.\m$ is the identity or returns a new object with the
same fields of the carrier, or with the two fields storing a same object, etc.

Therefore, in order to specify methods' types that are more informative than 
standard types,  we decided to use a further map -- 
the \emph{behavioural class table}, noted $\cct$.  The types used in the
$\cct$ are a variation of the above flattened types because we completely 
specify the tree structure of the object ($\cct$ is a global map; it is not
a vector of maps). These types are called 
\emph{structured types} and are ranged over by $\rho$, $\rho'$, $\cdots$. 
For example 
\[
(a[\f_1: (b[\g : \Int], \D), \f_2: (c[\g: \Int], \D)], \C)
\]
is an object of class $\C$ whose symbolic name is $a$ and that stores two 
different objects of class $\D$ in the fields. There is a simple 
way to transform a symbolic name and an environment into a structured type
and conversely, to get an environment out of a structured type. We 
call these functions $\constr{\Gamma,a}$ and $\destr{\rho}$, respectively,
and we leave their definitions as an exercise.

Let us discuss two examples of method types in the behavioural class table:
\begin{itemize}
\item
$\C.\m$ is the identity; hence it returns the carrier and the type also specifies 
that the carrier has not been modified. The method type is
\[
\cct(\C.\m) = (X, t, b) \rightarrow 
\langle X, X, \lam \rangle \; .
\] 
We notice that the type uses \emph{variable names}, ranged over by $X, Y, \cdots$, when the structure of
the argument is not relevant. Additionally, the arguments of $\C.\m$ are three:
the first element is the structured type of the carrier,
the second and the third arguments are two symbolic names. The name  
$t$ is the thread that performed the invocation and $b$ is the last 
object name whose lock has been acquired by $t$. 
These two informations are used by the analyser to build the right 
dependencies between callers and callees and appear in the lam $\lam$
of the return type. 

In the above method type, the carrier is addressed by $X$. This means that 
the symbolic name of the carrier is not used in the dependencies of $\lam$.
When this name is used, we write  $\cct(\C.\m)$ as 
{\footnotesize
\[
((a[\f_1:X, \f_2:Y],\C), t, b) \rightarrow 
\langle (a[\f_1:X, \f_2:Y],\C), \; (a[\f_1:X, \f_2:Y],\C), \; \lam \rangle \; ,
\] 
}
which binds the occurrences of $a$ in the return type.

\item
$\C.{\tt p}$ is the constructor of the class $\C$ that returns the carrier where 
the two fields have been initialised with the same new object of class $\D$
(we assume $\D$ has no field and we shorten $c[~]$ into $c$).
In this case, $\cct(\C.{\tt p})$ is
{\footnotesize
\[
\begin{array}{l}
((a[\f_1:X, \f_2:Y],\C), t, b) \rightarrow 
\\
\qquad \qquad \nulam{c}
\langle (a[\f_1:(c,\D), \f_2:(c,\D)],\C), \;
(a[\f_1:(c,\D), \f_2:(c,\D)],\C), \; \lam \rangle
\end{array}
\] 
}
The relevant part of the return type is $\nulam{c}$ part. This part
specifies that the name $c$ is \emph{new}, namely it does not occur in 
the arguments $(a[\f_1:X, \f_2:Y],\C), t, b$.
\end{itemize} 
The last concept we need for presenting the new rule for {\tt invokevirtual} is that of \emph{instance} of a method type. An instance of 
$\cct(\C.{\tt p})$ above when the arguments are 
$(a'[\f_1:\top, \f_2:\top],\C), t', b')$ (e.g.~$a'$ has been created without initialising the fields) is
{\footnotesize
\[
\langle (a'[\f_1:(c',\D), \f_2:(c',\D)],\C), \;
(a'[\f_1:(c',\D), \f_2:(c',\D)],\C), \; \lam\subst{a',t',b',c'}{a,t,b,c} \rangle \; .
\] 
}
This term will be written $\cct(\C.{\tt p})((a'[\f_1:\top, \f_2:\top],\C), t', b')(c')$.

The type rule for a method $\C.\m$ that updates the carrier 
(it is a constructor), has no argument and returns the updated carrier 
by creating one object is 
\[
\mathrule{t-invk}{
	\begin{array}{c}  
    P[i]= \invokevirtual~{\C.\m}~(~)\qquad
	i+1 \in \dom{P} \\
	 \of{S}{i} = a  \cdot S'
	\quad
	\typeof{\of{\Gamma}{i}}{a} = \C
	\\
	\recty = \constr{\of{\Gamma}{i}, a} 
	\quad
	b = \name{i}
	\quad
	\cct(\C.\m) (\recty,t, \ztop{\of{Z}{i}})(b) = \langle 
	\recty',
		\recty'', \lam \rangle
	\\
	\of{\Gamma}{i+1} = \of{\Gamma}{i} + \destr{\recty'} +	\destr{\recty''} 	\quad		
	\of{S}{i+1} =  \root{\recty'} \cdot S'
	\quad
    \of{F}{i+1}= \of{F}{i}
    \quad     \of{Z}{i+1} = \of{Z}{i}
	\end{array}
	}{ 
	   \cct, \Gamma, F,S,Z,i \vdash_t P: 
       \C.\m(\recty,t,\ztop{\of{Z}{i}})\rightarrow 
        \recty'
}
\]
The new part of this rule is the third line in the premise. 
In particular, in order to compute the instance of $\cct(\C.m)$, we 
construct $\constr{\of{\Gamma}{i}, a}$. The instance of the return type
$\langle \recty',\recty'', \lam \rangle$ 
is used to update $\of{\Gamma}{i}$ (in this case $\recty' = \recty''$,
therefore $\destr{\recty'}= \destr{\recty''}$). The function $\root{\recty}$
returns the root of the structured type $\recty$.

\paragraph{Concurrent methods.}
We finally discuss methods that are concurrent. Let $\C.\m$ be a method 
that creates a new thread, say $t'$ (and returns it). Since $t'$ runs in 
parallel with  the current thread, say $t$, the \emph{conjunctive} effects 
of  $t$ and $t'$
must be analysed by our tool (the second stage of our technique). 
To delegate the analyser to check the consistency of these 
conjunctive effects,  the 
type system must record the threads that are created. To this aim, we
 extend our judgments with a set 
collecting such thread names. However, this set may be infinite (when the 
method is recursive or iterative). In order to have a more precise
analysis, we distinguish the cases when the thread creation is executed 
once and those when the thread creation is executed several times. In the first case,
the analyser will spawn exactly one thread; in the second case the
analyser will spawn infinitely many threads (see the last part of the
paragraph ``\emph{Lams}''). 

As a consequence, our judgments
have two sets of thread names: $\tset$ for the names created once, $\rset$ 
for the names that will be spawned infinitely many times, each time with a 
fresh name, and they become
\[
		\cct,\Gamma, F,S,Z,\tset,\rset,i \vdash_t P: \lam
\]
We use the predicate ``$i$ \emph{is executed once}'' whenever the method
containing the instruction $i$ is not (mutual) recursive or the instruction
$i$ is not inside an iteration (this predicate can be easily computed in
our type system). 
The type rule for a method $\C.\m$ that creates two threads -- $t'$ 
executed once,  $t''$ spawned several times -- is 
\[
\mathrule{t-invk}{
	\begin{array}{c}  
    P[i]= \invokevirtual~{\C.\m}~(~)\qquad
	i+1 \in \dom{P} 
	\\
	\of{S}{i} = a \cdot S'
	\quad
	\typeof{\of{\Gamma}{i}}{a} = \C
	\\
	\recty = \constr{\of{\Gamma}{i}, a} 
	\quad
	t',t'' = \name{i}
	\quad
	\cct(\C.\m) (\recty,t, \ztop{\of{Z}{i}})(t',t'') = \langle 
	\recty',
	\{ t' \},\{ t''\},\recty'', \ell \rangle
	\\
	\of{\Gamma}{i+1} = \of{\Gamma}{i} + \destr{\recty'} + 
	\destr{\recty''} 
	\quad		
	\of{S}{i+1} =  \root{\recty'} \cdot S'
	\quad
    \of{F}{i+1}= \of{F}{i}
    \quad     \of{Z}{i+1} = \of{Z}{i}
    \\ 
     \of{\tset}{i+1}, \of{\rset}{i+1}  = \left\{ 
            \begin{array}{l@{\quad}l} 
       		\of{\tset}{i} \cup \{ t' \}, \; \of{\rset}{i} \cup \{ t'' \}
			& \text{if } i \text{ is executed once}
			\\
			\of{\tset}{i} , \; \of{\rset}{i} \cup \{ t', t'' \}  
			& \text{otherwise}
			\end{array} \right.
	\end{array}
	}{ 
		\cct,\Gamma, F,S,Z,\tset,\rset,i \vdash_t P: 
       \C.\m(\recty,t,\ztop{\of{Z}{i}})\rightarrow 
       \recty'
}
\]
In this case, the last premise defines the values of $\of{\tset}{i+1}$ and 
$\of{\rset}{i+1}$ according to the instruction $i$ is executed once or not.

\paragraph*{Lams.}
In our technique, the dependencies between symbolic names are expressed by means of \emph{lams}~\cite{GKL2014}, 
noted $\lam$, whose syntax is 
\[
\begin{array}{rl}
\lam \, ::= & \zero \quad | \quad \tpair{t}{a}{b} 
\quad | \quad 
\C.\m(\overline{\recty} 
) \rightarrow \recty' \quad | \quad \nulam{a} \lam 
\quad | \quad
\lam\, \&\, \lam
  \quad | \quad \lam + \lam 
\end{array}
\]
The term $\zero$ is the empty type; $\tpair{t}{a}{b}$ specifies a 
dependency between the object $a$ and the object $b$ 
that has been created by the thread $t$. The term 
$\C.\m(\overline{\recty}) \rightarrow \recty'$
defines the invocation of $\C.\m$  with arguments $\overline{\recty}$ and with returned  type $\recty'$. The argument sequence $\overline{\recty}$ 
has always at least three elements 
in our case: the first element is the carrier,
while the last two elements  are  
 the thread that performed the invocation and the last 
object name whose lock has been acquired by it. 
The 
operation $\nulam{a} \lam$ creates a new name $a$ whose scope
is the type $\lam$; the 
operations $\lam \, \&\, \lam'$ and $\lam + \lam'$ define
the conjunction and disjunction of the dependencies in $\lam$ and 
$\lam'$, respectively.
The operators $+$ and $\&$ are associative and commutative. 

A \emph{ lam program} is a pair $\bigl(\LAM, \ell \bigr)$, 
where $\LAM$ is a \emph{finite set} of \emph{function definitions}
\[
\C.\m(\overline{\recty}) \rightarrow \recty' \; = \; \ell_{\C.\m}
\]
with $\ell_{\C.\m}$ 
being the \emph{body} of $\C.\m$,
 and $\ell$ is the
\emph{main lam}. We notice that the type $\recty'$ is considered an 
argument of the lam function as well. When $\recty' = {\it void}$, 
the function definitions are shortened into $\C.\m(\overline{\recty}) = 
\ell_{\C.\m}$ and the invocations into $\C.\m(\overline{\recty})$.

As an example, 
the lams of the {\tt Network}'s code in 
Figure~\ref{fig:buildNetwork} is reported in 
Figure~\ref{fig:buildNetworkTypes} (lams have been simplified for easing the readability).
\begin{figure}
{\scriptsize
\begin{lstlisting}
Main(this,t,u) = ($\nu$ x,y)( Object.init(x,t,u)->x + Object.init(y,t,u)->y 
                  + buildNetwork(this,_,x,y,t,u) )

takeForks(this,x,y,t,u) = (u,x)$_{\tt t}$ $\&$ (x,y)$_{\tt t}$

buildNetwork(this,_,x,y,t,u) = ($\nu$ z,t1,u1)(  
      takeForks(this,x,y,t,u) 
      + Object.init(z,t,u) -> z  
      + Network$\$$1.init(t1[this$\$$0:$\top$,val$\$$x:$\top$,val$\$$z:$\top$],this,x,z,t,z) -> t1[this$\$$0:this,val$\$$x:x,val$\$$z:z] 
      + Network$\$$1.run(t1[this$\$$0:this,val$\$$x:x,val$\$$z:z],t1,u1) $\&$ buildNetwork(this,_,z,y,t,u) 
  
Object.init(this,t,u) -> this =   0

Network$\$$1.init(this[this$\$$0:X,val$\$$x:Y,val$\$$z:Z],x1,x2,x3,t,u) -> this[this$\$$0:x1,val$\$$x:x2,val$\$$z:x3] =   0

Network$\$$1.run(this[this$\$$0:x1,val$\$$x:x2,val$\$$z:x3],t,u) =   takeForks(x1,x2,x3,t,u)
\end{lstlisting}
}
\vspace{-.4cm}
\caption{{\tt Network}'s lams (the \mbox{\tt \_} is a place holder for an integer)}
\label{fig:buildNetworkTypes}
\vspace{-.3cm}
\end{figure}
We discuss the methods {\tt takeForks} and {\tt buildNetwork}.  The 
method {\tt takeForks} has arguments {\tt this}, {\tt x}, {\tt y}, {\tt t} 
and {\tt u}, where {\tt t} and {\tt u} are as discussed above. This method
acquires the locks of {\tt x} and {\tt y} in order; therefore its
lam is quite simple:  there is a dependency between {\tt u} and {\tt x} and
a dependency between {\tt x} and {\tt y}, namely {\tt (u,x)$_{\tt t}$ $\&$ (x,y)$_{\tt t}$}.
The lam of {\tt buildNetwork} is more complex. The first line corresponds to
the
{\tt then}-branch
(lines 0-10), namely the invocation to {\tt takeForks}. The other lines 
correspond to the {\tt else}-branch. Here we have the creation of the 
object {\tt z} and the invocation of the corresponding constructor (second line of the body of the lam function and line 17 of the bytecode), the
invocation to the constructor of {\tt Network}, that is called {\tt Network\$1},
which returns a new thread that we call {\tt t1}. The last line of the lam 
of {\tt buildNetwork} contains the invocation of {\tt t1.start} and 
the recursive invocation to {\tt buildNetwork}. These invocations are
 \emph{in conjunction} because
they are in parallel.

We conclude with a remark about the dependencies specified the 
the judgment 
\[
		\cct,\Gamma, F,S,Z,\tset,\rset,i \vdash_t P: \lam \; .
\]
In our system, these dependencies are actually those defined by $\lam$ 
\emph{and} those defined by
$\of{Z}{i}$, $\of{\tset}{i}$, and $\of{\rset}{i}$. In particular,
let $\of{Z}{i} = a \cdot a'$, $\of{\tset}{i} = \{ t' \}$, and $\of{\rset}{i} =
\{ t''\}$. Then the dependencies of the instruction $i$ are
($\constr{\of{\Gamma}{i}, t'} = (t'[\overline{\f : \recty'}],\C)$ and
$\constr{\of{\Gamma}{i}, t''} = (t''[\overline{\f : \recty''}],\C)$):
\[
\lam \; \& \; \tpair{t}{a'}{a} 
\; \& \; 
\C.{\tt run}( (t'[\overline{\f : \recty'}], \C), t', \lock{t'}) 
\; \& \; 
{\tt RUN}(t''[\overline{\f : \recty''}],\C)
\]
where $\C$ is a subclass of {\tt Thread}, $\lock{t'}$ is a (fake) name 
associated to $t'$ and representing a default object locked by $t'$, and 
{\tt RUN} is a lam function defined by 
\[
\begin{array}{rl}
{\tt RUN}(a[\overline{\f : \recty}], \set{\C}) = &
\C.{\tt run}( (a[\overline{\f : \recty}], \set{\C}), a, \lock{a}) \; 
\& \; \nulam{a'} \, {\tt RUN}(a'[\overline{\f : \recty}], \set{\C})
\end{array}
\]
The difference between $\tset$ and $\rset$ is exactly the fact that {\tt RUN} is
recursive. This means that every name in $\rset$ corresponds to the parallel
composition of infinitely many threads with different 
root names. The analyser in Section~\ref{sec:algorithm} verifies 
whether this composition is consistent or not (with respect to deadlocks).

\section{More about typing and {\JaDA}}
\label{sec.moreandJaDA}
%!TEX root = LOPSTR2017.tex

The type system described in this paper has been prototyped. 
It also covers features such as 
constructors, arrays, exceptions, static members, interfaces, inheritance, recursive data types. The overall system is called {\JaDA}.
Here we overview two relevant extensions --
inheritance and recursive data types --, the details of these two extensions and 
the other ones can be found in Garcia's PhD thesis~\cite{Garcia2017}.

\paragraph{Inheritance.}
{\JVMLd} does not admit to derive classes from other classes.
As a consequence, when a method is invoked, it is possible to uniquely 
locate the method definition (the output of $\typeof{\of{\Gamma}{i}}{a}$
in Section~\ref{sec:analysis} \emph{is always a single element}.
Therefore we cannot type 
{\tt 
\begin{center}
{\footnotesize
\begin{tabular}{l}
\ \ C w ;
\ \{ \ if (z) w = new D ; else w = new E ;
\ \} \ w.foo() ;
\end{tabular}
}
\end{center}}
\noindent
which is a correct {\Java} program, assuming that $\D$ and $\E$ are subclasses of $\C$.
In this case, if $\D$ and $\E$ have different implementations of {\tt foo}, 
we do not know how the invocation {\tt w.foo()} will be dispatched at run-time. 
Our solution consists of relaxing 
the relation 
between consecutive environments $\of{\Gamma}{i}$ and $\of{\Gamma}{i+1}$ in such a 
way that the type of $\of{\Gamma}{i+1}({\tt w})$ may be the one of 
$\of{\Gamma}{i}({\tt w})$ \emph{plus a set of subclasses therein}. Henceforth, the lam 
corresponding to  the {\tt invokevirtual} 
of {\tt w.foo()} is 
$\sum_{\C' \in \typeof{\of{\Gamma}{i}}{{\tt w}}}
\C'.{\tt foo}({\tt w}, t, a)$, namely $\C.{\tt foo}({\tt w}, t, a) + 
\D.{\tt foo}({\tt w}, t, a) + \E.{\tt foo}({\tt w}, t, a)$.

\paragraph{Recursive types.}
Recursive types are managed by using finite representations. Object names of recursive
types are \emph{special names} indexed by {\tt \$}.
A flattened \emph{recursive} record type is built
by unfolding the recursive types (exactly) up to those nodes
containing a name of a class already present in the tree. Nodes inside the tree
are labelled by new names, nodes
in the leaves are labelled either (for non recursive 
types) with $\top$ or $\Int$ or 
with names already present in the environment or (for recursive 
types) with names 
subscribed by a {\tt \$}  that correspond to the nodes of the classes that are already present in the tree.
By construction, these structures are finite. 
For instance, if $\C$ is a class whose type
is $[ {\tt val}: {\tt Thread}, \, {\tt next}: \C]$ (a list of threads) then, in correspondence of a 
{\tt new \C} instruction, we produce an environment 
$
\att{r} \mapsto [{\tt val}: (a[ \, ], {\tt Thread}), \, {\tt next}: \att{r}] \; .
$

Lists like the foregoing one are managed in ad-hoc ways. In particular we can deliver 
a precise analysis as long as \emph{the nodes of the list are all equal}, otherwise we 
return false positives. We observe that this technique is more precise than one
would think. For instance, assume to create a list of threads, where the field {\tt val}
of each node contains a new thread. This list is created by an iteration and 
the instruction creating the thread is always the same -- say $i$. Hence, by definition 
of the function $\name{i}$, the nodes of the list always contain the same name and can be
represented as described above.
Finally, in order to have a sound analysis, we also modify our definition of circularity 
in lams.
In particular, if the types of $\att{r}$ and of $\att{r'}$ are the same,
a term like $\tpair{t}{\att{r}}{c} \& \tpair{t'}{c}{\att{r'}}$ is a circularity because
$\att{r}$ may be replaced by \emph{every name of the same type}, including $\att{r'}$.

\section{The analysis of circularities in lams}
\label{sec:algorithm}
%!TEX root = LOPSTR2017.tex
\label{sec:fixpoint}

Once behavioural types have been computed for the whole {\JVMLd} program, 
we can analyse the type of the \emph{main} method. The analysis uses
an extension of the algorithm defined in~\cite{GKL2014,KobayashiLaneve} that we 
discuss below. 

The semantics of lams is very simple: it amounts
to unfolding function invocations. The critical points are that (\emph{i}) 
every invocation may create new fresh names and (\emph{ii}) the
function definitions may be recursive. These two points imply that a
lam model may have infinite states, which makes any analysis nontrivial. 
It is worth to  recall that the states of lams are conjunctions ($\aand$)
of dependencies and function invocation (because types with disjunctions $+$
are modelled by sets of states with conjunctive dependencies). 
The results of~\cite{GKL2014,KobayashiLaneve} allow us to reduce the
analysis to \emph{finite} models, \emph{i.e.}~finite disjunctions of 
finite conjunctions of dependencies. In turn, this finiteness makes 
possible to decide the 
presence of a so-called \emph{circularities}, namely terms such as $\tpair{t}{a}{b} \aand
\tpair{t'}{b}{a}$. 

In~\cite{GKL2014,KobayashiLaneve}, the
dependencies are not indexed by thread names:
here we use more informative dependencies in order to cope with {\Java} 
reentrant locks. In particular $\tpair{t}{a}{b} \aand \tpair{t}{b}{a}$ is not 
a circularity and, when $t \neq t'$, we carefully separate it from 
$\tpair{t}{a}{b} \aand \tpair{t'}{b}{a}$.
Because of this extension, we have modified the definitions of \emph{transitive 
closure} and of \emph{projecting-out fresh names}, which are basic notions in the 
algorithm of~\cite{GKL2014,KobayashiLaneve}.
Let $t \neq t'$ and let $\checkmark$ be a special object name. 
This symbol
$\checkmark$ is a special thread name indicating that the dependency 
is due to the contributions of two or more threads.
Let also $\lam$ be a conjunction $\aand$ of dependencies: 
\begin{itemize}
\item
the \emph{transitive
closure} of $\lam$, noted 
$\lam^+$, is the least conjunction that contains $\lam$ and such that
if $\tpair{t}{a}{b} \aand \tpair{t'}{b}{c}$ is a subterm of $\lam^+$ 
then either
(\emph{i}) $\tpair{\checkmark}{a}{c}$ is a subterm of $\lam^+$, if 
$t \neq t'$, or
(\emph{ii}) $\tpair{t}{a}{c}$ is a subterm of $\lam^+$, if $t = t'$;
\item
$\lam$ has a \emph{circularity} if 
there is $a$ such that $\tpair{\checkmark}{a}{a}$ is a subterm of  $\lam^+$. 
\end{itemize}
For example 
$\lam = \tpair{t}{a}{b} \aand \tpair{t}{b}{a}\aand \tpair{t'}{b}{c}$ has no
circularity because $\lam^+ = \tpair{t}{a}{b}\aand$ $\tpair{t}{b}{a} \aand 
\tpair{t}{a}{a} \aand \tpair{t}{b}{b}\aand \tpair{t'}{b}{c}\aand
\tpair{\checkmark}{a}{c}$ 
does not contain any pair $\tpair{\checkmark}{a}{a}$. 

As regards \emph{projecting-out fresh names}, when a lam $\ell$ contains a 
function invocation, our algorithm  replaces the invocation with the 
corresponding instance of its body \emph{where new names are replaced by fresh names}. For example, if $\ell =
\tpair{t}{a}{b}\aand f(a,b)$ and $f(x,y) = \nulam{z,t'}(\tpair{t'}{x}{y} \aand \tpair{t'}{x}{z})$ then 
we obtain the term $\ell' = \tpair{t}{a}{b}\aand \tpair{t'}{a}{b} \aand \tpair{t'}{a}{z'}$ -- where $z'$ is a fresh object name --,
which is equal to its transitive closure. We notice that $\ell'$ may be simplified: 
(\emph{i}) the dependency 
$\tpair{t'}{a}{z'}$ will never be involved in a circular dependency because $z'$ is
fresh in $t'$ and is unknown elsewhere; therefore it may be dropped;
(\emph{ii}) the dependency $\tpair{t'}{a}{b}$ is important, however the name 
$t'$ is not: we just need to separate it from the other (old thread) names. 
Therefore we replace $\tpair{t'}{a}{b}$ with $\tpair{\bullet}{a}{b}$, where $\bullet$
is a special thread name. The lam $\tpair{t}{a}{b}\aand \tpair{\bullet}{a}{b}$ is the
output of the projecting-out operation.

The algorithm we use is  the following. Let $\bigl(\LAM, \ell \bigr)$
be a \emph{lam program} and let $\LAM_0$ be the set of function definitions 
similar to $\LAM$ but with bodies $\zero$.

\begin{description}

\item[step 1:] $i = 0$;

\item[step 2:] 
compute $\LAM_{i+1}$: for every body $\ell_f$ of a lam function $f$ in $\LAM$
compute the new body ${\ell_f}^{(i+1)}$ as follows:
\begin{description}
\item[a.] replace bound names with fresh names;
\item[b.] replace function invocation in $\ell_f$ with their meaning in $\LAM_i$;
\item[c.] compute the transitive closure of the resulting lam and let ${\ell_f}^+$ be the 
new lam;
\item[d.] project out the fresh names in ${\ell_f}^+$ and let ${\ell_f}^{(i+1)}$ be the resulting lam.
\end{description}

\item[step 3:] if $\LAM_{i+1} \neq \LAM_i$ then $i = i+1$ and goto step 2, else
exit.
\end{description}
The above algorithm terminates and let $n$ be the least integer such that $\LAM_n = \LAM_{n+1}$.
It turns out that $\ell$ will display a circularity (by evaluating its function
invocations according to $\LAM$, which may be recursive) if and only if $\ell$ 
displays a circularity when the function invocations 
are evaluated with their definitions in $\LAM_n$ (which are not recursive). 

The proof of soundness of our type system is represented by a subject reduction 
theorem expressing that, if a {\JVM} configuration ${\it cn}$ has lam $\ell$
and \emph{cn} reduces to a configuration ${\it cn}'$ then (\emph{i}) ${\it cn}'$ is also 
well-typed and (\emph{ii}) if $\ell'$ is the type of ${\it cn}'$ then, if a 
circularity occurs in $\ell'$ then a circularity is present in $\ell$, as 
well.

\section{Assessment of {\JaDA}}
\label{sec.JaDA}
%!TEX root = LOPSTR2017.tex

Since {\JaDA} covers many features of {\Java}, it has been possible to 
deliver an initial assessment of it with respect to existing deadlock 
analysis tools. In particular, we have considered tools using different
techniques
{\tt Chord} for static analysis~\cite{naikPSG09},
{\tt Sherlock} for dynamic analysis~\cite{eslamimehrP14}, and 
{\tt GoodLock} for hybrid analysis~\cite{bensalemH05}.
We have also considered a commercial tool, {\tt ThreadSafe}~\footnote{http://www.contemplateltd.com/threadsafe}~\cite{threadSafe}.
Out of these tools, we were able to install and effectively test 
only two of them: {\tt Chord} and {\tt ThreadSafe}; the results corresponding 
to {\tt GoodLock} and {\tt Sherlock} come from \cite{eslamimehrP14}.
We also had problems in testing {\tt Chord} with some of the examples in the 
benchmarks, perhaps due to some misconfigurations, that we were not able to 
solve because {\tt Chord} has 
been discontinued. 

\begin{table}[h!] 
\vspace*{-.3cm}
  \centering
  \caption{Comparison with different deadlock detection tools. The
  inner cells show the number of deadlocks detected by each tool. 
  The output labelled ``(*)'' are related to modified versions of the original programs: see the text.}
  \label{tab:comparisonone}
{\scriptsize  \hspace*{-.7cm}\begin{tabular}{|l|l|l|| l|l|l|l|l|}
  \hline & & &\multicolumn{2}{c|}{Static}& Hybrid & Dynamic & Commercial\\
  \hline
  benchmarks & LOC, \#Threads & deadlock & {\JaDA}[tm] & {\tt Chord}[tm] & {\tt GoodLock}[tm] & {\tt Sherlock}[tm] & {\tt ThreadSafe}[tm]\\
  \hline
  Sor & 1274, \quad 5 & yes & 1 \; [135s] & 1 \; [210s]& 7 \; [4s] & 1 \; [39s] & 4 \; [435s]\\
  RayTracer(*) & 1292, \quad 5  & no & $0$ \; [155s] & 0 \; [223s]& 8 \; [2s] & 2 \; [30s]& 0 \; [502s]\\
  MolDyn (*) & 1351, \quad 5 & no & $0$ \; [110s] & 0 \; [191s] & 6 \; [5s] & 1 \; [49s]& 0 \; [423s]\\
  MonteCarlo (*) & 3619, \quad 4 & no & $0$ \; [231s]& 0 \; [342s]& 23 \; [5s] & 2 \; [102s]  & 0 \; [821s] \\
   \hline
  BuildNetworkN & 40, \quad N+1 & yes & 3 \; [8s] & 0 \; [50s]&  &  & 0 \; [50s]\\
  PhilosophersN & 60, \quad N+1  &  yes & 3 \; [12s] & 0 \; [51s] &  &  & 0 \; [51s]\\
  ThreadArraysN & 23, \quad N+1 &  yes & 1 \; [6s] & 1 \; [40s]&  &  & 1 \; [40s]\\
  ThreadArraysJoinsN & 37, \quad N+1 &  yes & 1 \; [6s] & 1 \; [41s] &  &  & 0 \; [41s]\\
  \hline
  ScalaSimpleDeadlock & 39, \quad 2 &  yes & 1 \; [3s]&  &  &  & \\
  ScalaPhilosophersN & 62, \quad N+1 &  yes & 3 \; [4s] &  &  &  & \\
  \hline
  \end{tabular}}
\vspace*{-.3cm}
\end{table}

We have analysed a number of programs that exhibit a variety of sharing 
patterns. 
The source of all benchmarks in Table~\ref{tab:comparisonone} 
is available either at \cite{eslamimehrP14,naikPSG09} or in the 
{\JaDA}{\tt-deadlocks} 
repository\footnote{https://github.com/abelunibo/Java-Deadlocks}.
Since the current release of {\JaDA} does not completely cover {\JVM},
in order to gain preliminary experience, we modified the Java libraries 
and the multithreaded server programs of RayTracer, MolDyn and MonteCarlo
(labelled with ``(*)'' in the Table~\ref{tab:comparisonone})
and implemented them in our system. This required little programming
overhead; in particular, we removed volatile variables, 
avoided the use of 
{\tt Runnable} interfaces for
creating threads,  and reduced the invocations of native methods 
involved in I/O operations. For every program, we give the lines of code (LOC), 
the number \#Threads of threads explicitly created (in the second and third block this number depends on
the argument N). We also state whether the program under examination has a deadlock or not
and the time in seconds (tm) each tool took to perform the analysis. The times for
{\tt GoodLock} and {\tt Sherlock} were taken from the literature~\cite{eslamimehrP14}.

Here are our remarks. 
The first block of programs belongs to a well known group  used as 
benchmarks for several {\Java} analysis tools; the second block corresponds 
to examples designed to test {\JaDA} 
against complex deadlock scenarios.
First of all {\JaDA} is the unique tool that never returns 
false positives or false negatives. {\tt Chord} and {\tt ThreadSafe} are unsound
because they return false negative (see the second block). The execution time
of the tools are similar ({\JaDA} appears more efficient), except for 
{\tt GoodLock} and {\tt Sherlock}, which appear however much less precise (they 
return a lot of false positives). As regards the second block, we observe that 
{\JaDA} returns few deadlocks, which do not depend from N. This is because our
analysis is symbolic and does not  consider numeric values (most of the deadlock are
considered ``to be similar''). 				
					
The third group reports the analysis of two examples of {\tt Scala} 
programs~\cite{Scala} (the {\tt Scala}
compiler $2.11$ produces {\Java} bytecode). To the best 
of our knowledge, there is no static 
deadlock analysis tool for {\tt Scala} (for this reason the entries
corresponding to the other tools are empty).

We have also analyzed the whole {\Java} library. The overall analysis took 5 hours and 40 min. We have considered as entry points the public static parameterless methods
and we have run the analyzer with the following limitations: native codes are
not analyzed (their behavioural type is $\zero$) and concurrency dependencies caused 
by {\tt wait}/{\tt notify} patterns are not verified.
The analysis has not reported any deadlock.

\section{Related work and Conclusions}
\label{sec:conclusions}
%!TEX root = LOPSTR2017.tex

We do not have space to discuss in detail the related work; therefore we
focus on the tools used in the assessment of Section~\ref{sec.JaDA} and
their theories.
{\tt ThreadSafe} uses a data-flow analysis that constructs
an execution flow graph and searches for cycles within this graph. 
Some heuristics are used to remove likely false positives.
No alias analysis to resolve object identity across method calls is attempted.
This analysis is performed in {\tt Chord}~\cite{eslamimehrP14,naikPSG09}, which 
can detect re-entrance on restricted cases, such as when lock expressions 
concern local variables (it is not 
possible to use fields). 
{\tt GoodLock}~\cite{bensalemH05} and its refinement  
{\tt Sherlock}~\cite{eslamimehrP14} use a theory that is based on monitors. 
Therefore the technique is a runtime technique that
tags each segment of the program reached by the 
execution flow and specifies the exact order of lock 
acquisitions. Thereafter, these segments are analyzed
for detecting potential deadlocks that might occur because
of different scheduler choices (than the current one). This kind of technique is 
partial because one might overlook sensible patterns of methods' arguments 
(\emph{cf.} {\tt BuildNetwork}, for instance). We are not aware of other
static analysers for the deadlock detection of {\Java}. A powerful static tool 
is {\tt SACO}~\cite{Antonio2013}, which has been developed for {\tt ABS}, an 
object-oriented language with a
concurrent model different from {\Java}. A comparison between {\tt SACO} and
a tool using a technique similar to the one in this paper can be found in~\cite{giachinoLL15}.

In this paper we have defined a new technique for detecting deadlocks in {\Java} programs 
by analysing the {\Java} intermediate language {\JVML}. 
The technique has been specified by focusing on a subset of {\JVML} featuring
thread creations and synchronizations, called {\JVMLd}.
We have also developed a prototype, called {\JaDA}, which also covers 
complex features of {\Java}, such as static members, arrays, 
recursive data types, 
exception handling, inheritance and dynamic dispatch. 
These extensions have made possible to 
deliver an initial assessment of {\JaDA} with respect to existing deadlock 
analysis tools for {\Java}.

Our future work includes the analysis of features of {\Java} that 
have not yet been studied. 
One relevant feature is thread coordination, which is expressed by the methods {\tt wait}, {\tt notify} and 
{\tt notifyAll}. 
Another extension addresses \emph{native methods}, namely methods
that are not implemented within the language and that 
are used when it is necessary to
interact with the Operating System or for meta-programming purposes.
Our current solution is to
manually insert in the {\cct} the behavioural types of native methods.
We are investigating testing mechanisms that may help in writing the 
types of such methods.

\bibliography{biblio}

\begin{thebibliography}{10}

\bibitem{threadSafe}
Robert Atkey and Donald Sannella.
\newblock Threadsafe: Static analysis for {Java} concurrency.
\newblock {\em {ECEASST}}, 72, 2015.
\newblock URL: \url{http://journal.ub.tu-berlin.de/eceasst/article/view/1025}.

\bibitem{bensalemH05}
Saddek Bensalem and Klaus Havelund.
\newblock Dynamic deadlock analysis of multi-threaded programs.
\newblock In {\em in Hardware and Software Verification and Testing}, volume
  3875 of {\em Lecture Notes in Computer Science}, pages 208--223. Springer,
  2005.

\bibitem{eslamimehrP14}
Mahdi Eslamimehr and Jens Palsberg.
\newblock Sherlock: scalable deadlock detection for concurrent programs.
\newblock In {\em Proceedings of the 22nd International Symposium on
  Foundations of Software Engineering (FSE-22)}, pages 353--365. {ACM}, 2014.

\bibitem{Antonio2013}
Antonio Flores-Montoya, Elvira Albert, and Samir Genaim.
\newblock May-happen-in-parallel based deadlock analysis for concurrent
  objects.
\newblock In {\em Proc. FORTE/FMOODS 2013}, volume 7892 of {\em Lecture Notes
  in Computer Science}, pages 273--288. Springer, 2013.

\bibitem{Garcia2017}
Abel Garcia.
\newblock {\em Static analysis of concurrent programs based on behavioral type
  systems}.
\newblock PhD thesis, {School in Computer Science and Engineering}, 2017.
\newblock {Available at {\tt JaDA.cs.unibo.it}}.

\bibitem{GKL2014}
Elena Giachino, Naoki Kobayashi, and Cosimo Laneve.
\newblock Deadlock analysis of unbounded process networks.
\newblock In {\em Proceedings of 25th International Conference on Concurrency
  Theory {CONCUR} 2014}, volume 8704 of {\em Lecture Notes in Computer
  Science}, pages 63--77. Springer, 2014.

\bibitem{GiachinoL14}
Elena Giachino and Cosimo Laneve.
\newblock Deadlock detection in linear recursive programs.
\newblock In {\em 14th Int.~School on Formal Methods for the Design of
  Computer, Communication, and Software Systems ({SFM} 2014)}, volume 8483 of
  {\em Lecture Notes in Computer Science}, pages 26--64. Springer, 2014.

\bibitem{giachinoLL15}
Elena Giachino, Cosimo Laneve, and Michael Lienhardt.
\newblock A framework for deadlock detection in core {ABS}.
\newblock {\em Software and Systems Modeling}, 15(4):1013--1048, 2016.

\bibitem{KobayashiLaneve}
Naoki Kobayashi and Cosimo Laneve.
\newblock Deadlock analysis of unbounded process networks.
\newblock {\em Inf. Comput.}, 252:48--70, 2017.

\bibitem{naikPSG09}
Mayur Naik, Chang{-}Seo Park, Koushik Sen, and David Gay.
\newblock Effective static deadlock detection.
\newblock In {\em 31st International Conference on Software Engineering ({ICSE}
  2009)}, pages 386--396. {ACM}, 2009.

\bibitem{Scala}
Martin Odersky and al.
\newblock An {O}verview of the {S}cala {P}rogramming {L}anguage.
\newblock Technical Report IC/2004/64, EPFL, Lausanne, Switzerland, 2004.

\end{thebibliography}

\end{document}